\documentclass[10pt,aps,prl,twocolumn,showpacs,superscriptaddress,
longbibliography]{revtex4-1}

\usepackage{graphicx}
\usepackage{amsmath}
\usepackage{amssymb}
\usepackage{bm}
\graphicspath{{./FiguresPDF/}}
\usepackage{xcolor}
\usepackage[%
  colorlinks=true,
  urlcolor=blue,
  linkcolor=blue,
  citecolor=blue
]{hyperref}

\newcommand{\lkb}{Laboratoire Kastler Brossel, Coll\`ege de France, CNRS, ENS-PSL Research University, UPMC-Sorbonne Universit\'es, 11 place Marcelin Berthelot, 75005 Paris, France}
\begin{document}
\title{Stepwise Bose-Einstein condensation in a spinor gas}
%
%%Successive Bose-Einstein condensations in a spinor gas
%%Sequential Bose-Einstein condensation in a spinor gas}
%%
\author{C.\ Frapolli}
		\affiliation{\lkb}
		
\author{T.\ Zibold}
		\altaffiliation{Current address: Department of Physics, University of Basel, Klingelbergstrasse 82, 4056 Basel, Switzerland}
		\affiliation{\lkb}

\author{A.\ Invernizzi}
		\affiliation{\lkb}
		
\author{K.\ Jim\'enez-Garc\'ia}
		\affiliation{\lkb}

\author{J.\ Dalibard}
		\affiliation{\lkb}
		
\author{F.\ Gerbier}
		\affiliation{\lkb}
		
\date{\today}

\begin{abstract}
%
%We measure the thermodynamic phase diagram of antiferromagnetic spin $F=1$ bosons using ultracold Sodium atoms in a tight optical trap.
We observe multi-step condensation of sodium atoms with spin $F=1$, where the different Zeeman components $m_F=0,\pm 1$ condense sequentially as the temperature decreases. The precise sequence changes drastically depending on the magnetization $m_z$ and on the quadratic Zeeman energy $q$ (QZE) in an applied magnetic field. For large QZE, the overall structure of the phase diagram is the same as for an ideal spin 1 gas, although the precise locations of the phase boundaries are significantly shifted by interactions. For small QZE, antiferromagnetic interactions qualitatively change the phase diagram with respect to the ideal case, leading for instance to condensation in $m_F=\pm 1$, a phenomenon that cannot occur for an ideal gas with $q>0$.
\end{abstract}
\maketitle

Multi-component quantum fluids described by a vector or tensor order parameter are often richer than their scalar counterparts. Examples in condensed matter are superfluid $^3$He \cite{vollhardt3He} or some unconventional superconductors with spin-triplet Cooper pairing \cite{norman2011a}. In atomic physics, spinor Bose-Einstein condensates (BEC) with several Zeeman components $m_F$ inside a given hyperfine spin $F$ manifold can display non-trivial spin order at low temperatures \cite{ho1998a,ohmi1998a,stenger1998a,stamperkurn2013a}. The macroscopic population of the condensate enhances the role of small energy scales that are negligible for normal gases. This mechanism (sometimes termed \textit{Bose-enhanced magnetism} \cite{stamperkurn2013a}) highlights the deep connection between Bose-Einstein condensation and magnetism in bosonic gases, and raises the question of the stability of spin order against temperature.

In simple cases, magnetic order appears as soon as a BEC forms. Siggia and Ruckenstein \cite{siggia1980a} pointed out for two-component BECs \cite{siggia1980a} that a well-defined relative phase between the two components implies a macroscopic transverse spin. BEC and ferromagnetism then occur simultaneously, provided the relative populations can adjust freely. A recent experiment confirmed this scenario for bosons with spin-orbit coupling \cite{ji2014a}. This conclusion was later generalized to spin-$F$ bosons without \cite{yamada1982a} or with \textit{spin-independent} \cite{eisenberg2002a} interactions. These results indicate that without additional constraints, bosonic statistics favors ferromagnetism.

In atomic quantum gases with $F>1/2$, this type of ferromagnetism competes with spin-exchange interactions, which may favor other spin orders such as spin-nematics \cite{stamperkurn2013a}. %In particular, Van der Waals interactions between alkali atoms are not spin-independent. 
Spin-exchange collisions can redistribute populations among the Zeeman states \cite{chang2004a,schmaljohann2004a,kuwamoto2004a}, but are also invariant under spin rotations. The allowed redistribution processes are therefore those preserving the total spin, such as $2\times (m_F=0) \leftrightarrow (m_F=+1)+(m_F=-1)$.
%, in contrast with, \textit{e.g.}, mixtures of different atomic species. 
%Spin-exchange interactions they are invariant under spin rotations \cite{stamperkurn2013a}. 
For an isolated system driven to equilibrium only by binary collisions (in contrast with solid-state magnetic materials \cite{levy_en}), and where magnetic dipole-dipole interactions are negligible (in contrast with dipolar atoms \cite{pasquiou2012a}), the longitudinal magnetization $m_z$ is then a conserved quantity. This conservation law has deep consequences on the thermodynamic phase diagram.

The thermodynamics of spinor gases with conserved magnetization has been extensively studied theoretically using various assumptions and methods  \cite{isoshima2000a,zhang2004a,kao2006a,lang2014a,uchino2010a,phuc2011a,kawaguchi2012b}. A generic conclusion is that Bose-Einstein condensation occurs in steps, where BEC occurs first in one specific component and magnetic order appears at lower temperatures when two or more components condense. %To the best of our knowledge, this scenario has not been observed experimentally before. 
Natural questions are the number of steps that can be expected, and the nature of the magnetic phases realized at different temperatures. %For a non-interacting spin-1 BEC, only two steps are expected, while in the interacting case, either two or three steps are predicted in \ref{} for antiferromagnetic or ferromagnetic interactions.
%For spin-$F$ bosons, it seems natural to expect $2F+1$ steps,  

\begin{figure}[ht!!!]
\centering
\includegraphics[width=0.45\textwidth]{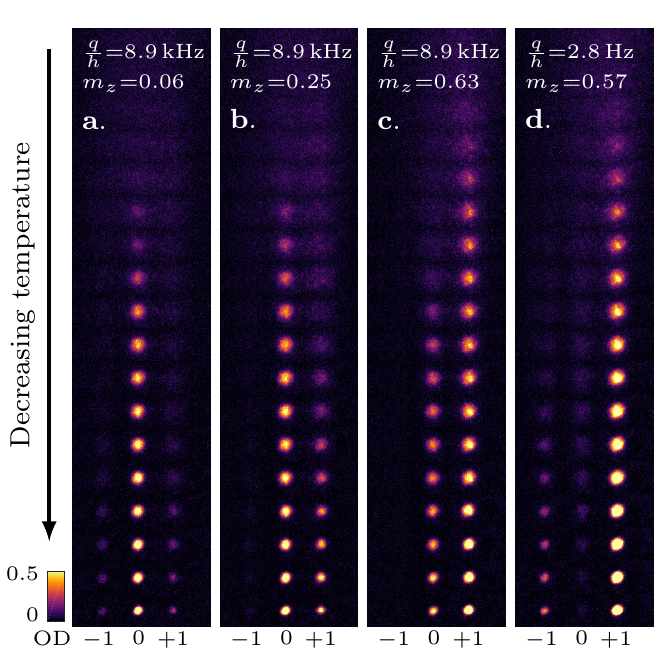}
\caption{ Illustration of stepwise Bose-Einstein condensation in antiferromagnetic spin 1 gases. Each column is formed by juxtaposing absorption images of spin distributions with monotonically decreasing temperature $T$ from top to bottom. The quadratic Zeeman energy $q$ and low-$T$ magnetization $m_z$ are indicated at the top of each column. \textbf{a.} Only $m_F=0$ condenses ($m_F=0,\pm1$ are the three Zeeman states). \textbf{b.} (\textbf{c.}) For low (high) magnetizations, $m_F=0$     ($m_F=+1$) condenses first followed by $m_F=+1$ ($m_F=0$). \textbf{d.} For small $q$ and high $m_z$, $m_F=+1$ condenses first followed by $m_F=-1$, while $m_F=0$ does not condense.} \label{fig1}
\end{figure}

In this Letter, we report on the observation of multi-step condensation in an antiferromagnetic $F=1$ condensate of sodium atoms. Fig.\,\ref{fig1} illustrates four situations that occur when lowering the temperature starting from a normal Bose gas. Without loss of generality, we focus in this work on the case of positive magnetization, given that the case of $m_z < 0$ can be deduced by symmetry. In all cases with $m_z\neq 0$, we find a sequence of transitions where different Zeeman components condense at different temperatures. Depending on the applied magnetic field $B$ and on the magnetization, we find either two or three condensation temperatures. The purpose of this paper is to explore this rich landscape of transitions in a bosonic spinor system and to elucidate the role of atomic interactions.

The present work is to the best of our knowledge the first comprehensive measurement of thermodynamic properties of spinor condensates with conserved magnetization. 
Previous experimental works exploring finite temperatures in spinor gases mostly studied spin dynamics in thermal gases \cite{pechkis2013a,he2015a,erhard2004a,naylor2016a}, or demonstrated cooling of a majority Zeeman component by selective evaporation of the minority components \cite{Olf2015a,naylor2015a}. The realization of dipolar spinor gases with free magnetization \cite{pasquiou2012a} was limited to the study of spin-polarized condensed phases in equilibrium due to dipolar relaxation. More recently, a gas of spin excitations in a spin-polarized ($m_z \approx 1$) ferromagnetic Bose-Einstein condensate was observed to equilibrate and even condense at sufficiently low temperatures \cite{fang2016a}.

%In contrast to the sparse experimental literature, thermodynamics of spinor gases has been extensively studied theoretically using various assumptions and methods. The ideal spin 1 gas has been discussed in Refs.\,\cite{isoshima2000a,kao2006a} for $q=0$, and  in Ref.\,\cite{lang2014a} for $q\neq 0$. The role of interactions has been explored using various approximations \cite{isoshima2000a,zhang2004a,uchino2010a,phuc2011a,kawaguchi2012b}. However, our experimental results can be fully understood only when taking into account three important features, the harmonic trap (which is crucial to stabilize an antiferromagnetic condensate in a single spatial mode \cite{yi2002a}), the QZE and the interactions.

%The magnetization acts here as an external control parameter independent of the externally applied magnetic field $B$, in contrast with dipolar atoms \cite{pasquiou2012a} or with solid-state magnetic materials \cite{levy_en}. The linear Zeeman effect thus becomes irrelevant for the equilibrium state, since it is proportional to a conserved quantity. The main role of $B$ is therefore to lower the energy of $m_F=0$ with respect to $m_F=\pm1$  by an amount $q\propto B^2$ (the quadratic Zeeman energy -- QZE).

Our experiments are performed with ultracold $^{23}$Na atoms confined in a crossed optical dipole trap (ODT). The longitudinal magnetization $m_z=(N_{+1}-N_{-1})/N$ acts as an external control parameter independent of the externally applied magnetic field $B$. Here, $N_{m_F}$ is the reduced population in Zeeman state $m_F$ and $N$ the total atom number. We vary $m_z$ between unmagnetized ($m_z\approx 0$) and fully magnetized samples ($m_z\approx 1$) using a preparation sequence performed far above $T_c$ \cite{jacob2012a,SM}. An applied magnetic field $B$ shifts the single-atom energy by $\Delta E_{m_F}=p m_F+q(m_F^2-1)$. The conservation of magnetization makes the linear Zeeman effect $\propto p$ irrelevant in the equilibrium state. The quadratic Zeeman energy (QZE), which lowers the energy of $m_F=0$ with respect to $m_F=\pm1$, is the relevant term, and is given by  $q= \alpha_q B^2$ with $\alpha_q/h \approx 277\,$Hz/G$^2$ for sodium atoms. 

The depth $V_0$ of the ODT determines the temperature $T$ and total atom number $N$ for a given $V_0$. We find that the magnetization $m_z$ also varies with $V_0$ (by up to $15\,$\%), a byproduct of evaporative cooling. Once a condensate forms in one of the Zeeman components, evaporation tends to eliminate preferentially atoms in the other Zeeman states. The evaporative cooling dynamics is very slow compared to the microscopic thermalization time on which the gas returns to thermal equilibrium. As a result, the kinetic equilibrium state for the quantum gases studied in this work is still determined by a magnetization-conserving Hamiltonian. Furthermore, the ODT is tight enough such that a condensate forms in the so-called single-mode regime \cite{yi2002a}, where the spatial shape of the condensate wavefunction is independent of the Zeeman state. In the following, we characterize our data for a given value of $q$ by an evaporation ``trajectory'' $(N,T,m_z)_{V_0}$, taking four experimental realizations for each point in the trajectory.

\begin{figure}[t]
\centering
\includegraphics[width=0.45\textwidth]{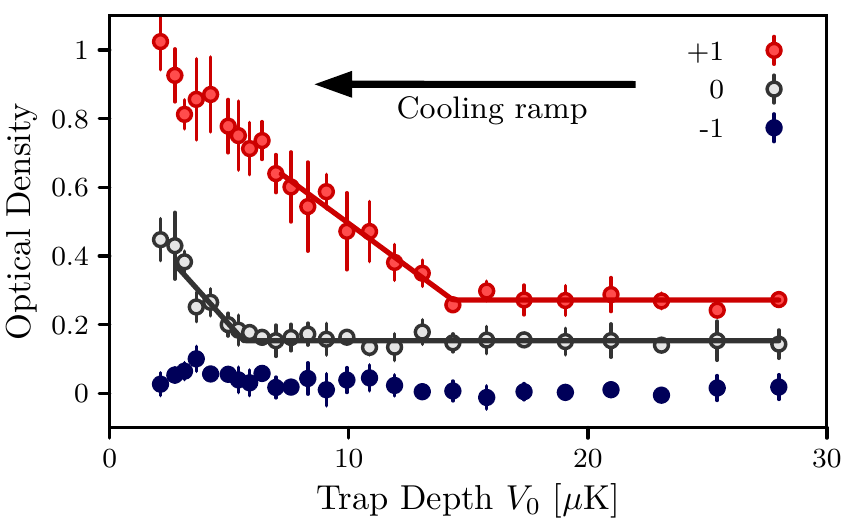}
\caption{Evolution of peak optical density with trap depth for a particular evaporation trajectory with $q/h\approx 69\,$Hz and $m_{z}\approx 0.3$ at the highest temperature. For these parameters, the $m_F=+1$ component condenses first (at a temperature $T_{c,1}\approx 1.8\,\mu$K), followed by the $m_F=0$ component (at a temperature $T_{c,2}\approx 560\,$nK). No condensate was detected in the $m_F=-1$ component. The curves for $m_F=+1$ and $m_F=0$ have been shifted vertically by 0.2 and 0.1 for clarity. The error bars denote statistical uncertainties at a 66\,\% confidence level. The solid lines indicate the piece-wise linear fits used to determine the critical trap depths.} \label{fig2}
\end{figure}

\begin{figure*}
\includegraphics[width=\textwidth]{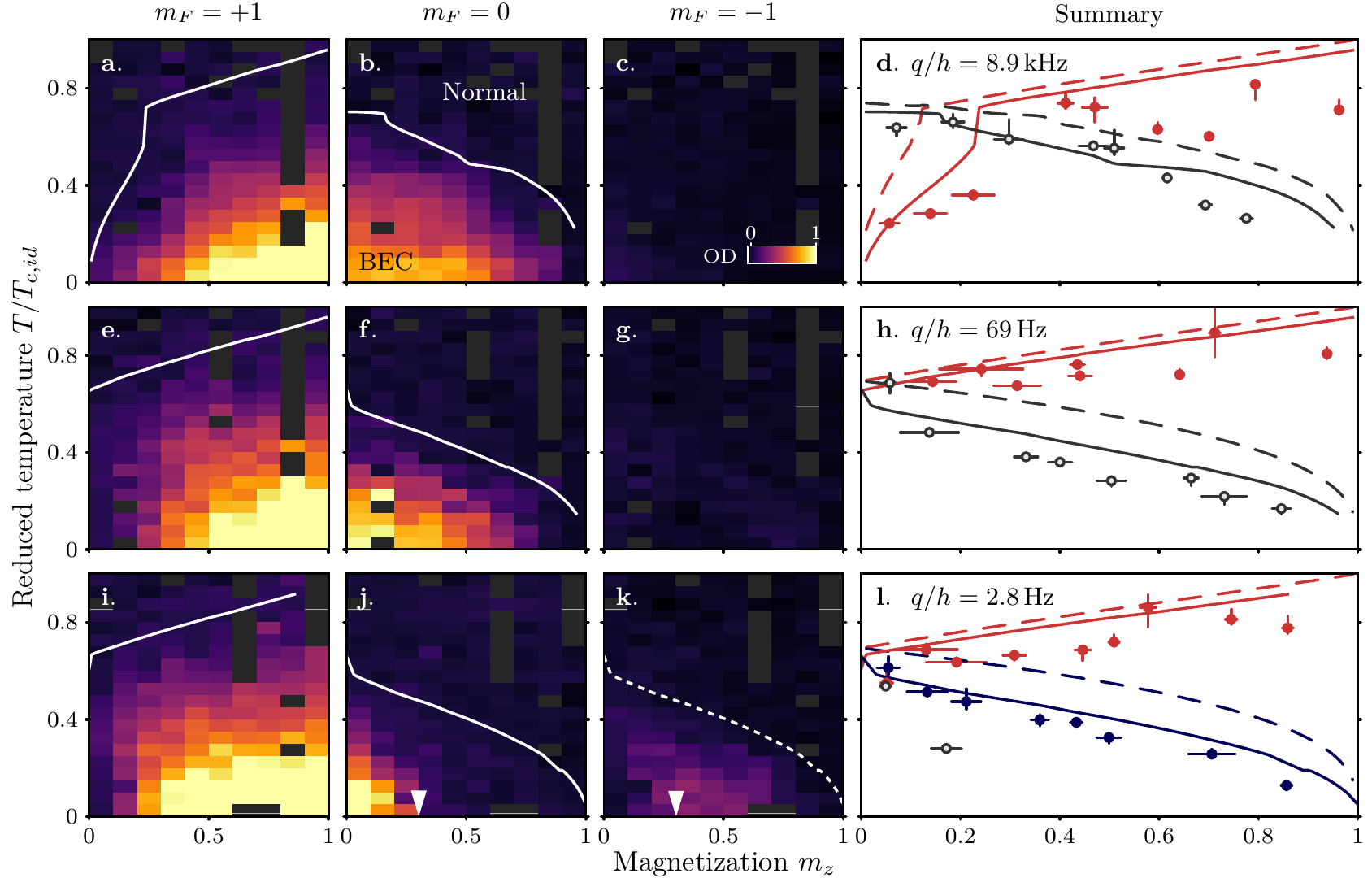}
\caption{Thermodynamic phase diagram of an antiferromagnetic spin $F=1$ Bose gas. The peak optical density of each Zeeman component is reported for the entire set of data at each value of the QZE -- $q/h=8.9\,$kHz (\textbf{a-c}), $69\,$Hz (\textbf{e-g}) and $q=2.8\,$Hz (\textbf{i-k}). The temperature is normalized to $T_{c,\textrm{id}}$, the critical temperature of a \textit{single-component} ideal Bose gas with the same number of atoms. The grayed areas indicate the absence of data in the corresponding regions. The right column (\textbf{d,h,l}) shows the measured critical temperatures of the $m_F=+1,0,-1$ Zeeman components (red, gray, and blue markers, respectively). The solid (dashed) lines are the predictions of a Hartree-Fock (HF) model with spin-independent interactions (ideal gas theory). The dotted line in $\textbf{k}$ shows the expected $T_{\rm c,2}$ where $m_F=0$ condenses according to the HF model.}
\label{fig3}
\end{figure*}

Absorption images as shown in Fig.\,\ref{fig1} are recorded after $3\,$ms of expansion in an applied magnetic field gradient \cite{SM}. We perform a fit to a bimodal distribution for each component to extract the temperature, the populations $N_{m_F}$, and the condensed fraction $f_{c,{m_F}}$ per component \cite{SM}. We found that low condensed fractions $< 5 \%$ are difficult to detect with the fit algorithm due to a combination of low signal-to-noise ratio and the complexity of fitting the three Zeeman components simultaneously. The signature of BEC, the appearance of a dense, narrow peak near the center of the atomic distribution, can instead be tracked by monitoring the peak optical density (OD) taken as a proxy for the condensed fraction \cite{trotzky2010a}. This procedure avoids relying on bimodal fits or other indirect analyses with uncontrolled systematic biases. 

Fig.\,\ref{fig2} shows such a measurement for a particular evaporation trajectory. The peak OD increases sharply when Bose-Einstein condensation is reached, demonstrating in this particular example a two-step condensation where $m_F=+1$ condenses first, followed by $m_F=0$. For a given evaporation trajectory, we identify the critical trap depth $V_{\rm 0,c}$ where condensation is reached by a piece-wise linear fit to the data, taking the intercept point as the experimentally determined $V_{\rm 0,c}$ (see Fig.\,\ref{fig2}). We interpolate numerically the atom number, magnetization and temperature to obtain the critical values $N_{\rm c}$, $T_{\rm c}$, $m_{z,{\rm c}}$ from $V_{\rm 0,c}$.

Fig.\,\ref{fig3} summarizes the results of this work. We show the peak optical density for each Zeeman component and each value of $q$ in a $(T-m_z)$ plane (Fig.\,\ref{fig3}\,a-c,\,e-g and\,i-k). In this plot, all data taken at a given QZE $q$ are binned with respect to magnetization and temperature. The domains where condensation occurs appear in light colors. For convenience, the temperature is scaled to the critical temperature of a \textit{single-component} ideal gas
$k_B T_{c,\textrm{id}}=\hbar\overline{\omega}[N/\zeta(3)]^{1/3}$, with $\overline{\omega}$ the geometric average of the trap frequencies and $\zeta$ the Riemann zeta function \cite{dalfovo1999a}. The same plot also shows the measured critical temperatures (Fig.\,\ref{fig3}\,d,\,h,\,l)\footnote{In one case, $m_F=0$ when $m_z \approx 0.3$ and $q/h=2.8\,$Hz, the lowest temperature images do show a condensed component but the critical temperature could not be extracted reliably from the fitting procedure due to sparse sampling. This particular point is not reported in Fig.\,\ref{fig3}l.}. The phenomenon of sequential condensation is always observed for $m_z \neq 0$, but the overall behavior changes drastically with $q$.

We first discuss the cases with largest QZE, $q/h\approx 8.9\,$kHz (Fig.\,\ref{fig3}\,a-d) and $q/h\approx 69\,$Hz (Fig.\,\ref{fig3}\,e-h). %The last case with small $q/h\approx3\,$Hz will be discussed later. 
For $q/h\approx 8.9\,$kHz and highly magnetized samples, the majority component $m_F=+1$ condenses first at a critical temperature $T_{c,1}$, followed by the $m_F=0$ component at a lower temperature $T_{c,2}$. For low magnetizations, the condensation sequence is reversed. For $q/h\approx 69\,$Hz, we observe only one sequence, a two-step condensation with $m_F=+1$ first and $m_F=0$ second.

This behavior can be understood qualitatively from ideal gas theory, taking the QZE and the conservation of magnetization into account \cite{lang2014a}. For ideal gases, BEC occurs when the chemical potential $\mu$ equals the energy of the lowest single-particle state \cite{dalfovo1999a}. The same criterion holds for a spin 1 gas with $\mu_0=\mu$ and $\mu_{\pm 1}=\mu \pm \lambda$, where $\lambda$ is a Lagrange multiplier enforcing the conservation of $m_z$. For $m_z =0$ ($\lambda=0$) and $q >0$, the QZE lowers the energy of $m_F=0$, which is therefore the first component to condense when $\mu=-q$. For $m_z > 0$, $\lambda$ is positive and increases with $m_z$. The energetic advantage of $m_F=0$ is in balance with the statistical trend favoring the most populated component $m_F=+1$. Eventually, this trend takes over at a ``critical'' value $m_z^\ast$ (where $\lambda=q$). For $m_z>m_z^\ast$, the $m_F=+1$ component condenses first.

Coexisting $m_F=0$ and $m_F=\pm 1$ components with a well-defined phase relation correspond to a non-zero transverse spin $\langle \hat{S}_x+i\hat{S}_y \rangle \neq 0$ (``transverse magnetized'' phase -- $M_\perp$). For large $q$, the condensate is reduced to an effective two-component system $m_F=0,+1$ with $m_F=-1$ mostly spectator. The case $m_z=m_z^\ast$ ($\mu_{0}=\mu_{+1}$) realizes the Siggia-Ruckenstein (S-R) scenario, where condensation and ferromagnetic behavior appear simultaneously. Away from that point, the S-R picture breaks down ($\mu_{0} \neq\mu_{+1}$) and sequential condensation takes place. 

Figure\,\ref{fig3}\,d-h show the critical temperatures and compare them to ideal gas theory. Although the general trends in the theory are the same as in the experiment, we observe a systematic shift of $T_{c,1}$ and $T_{c,2}$ towards lower temperatures, and an experimental ``critical'' $m_z^\ast \sim 0.3$ larger than the ideal gas prediction. The behavior for $q/h\approx 69\,$Hz (Fig.\,\ref{fig3}\,e-h) is qualitatively similar to the largest $q$ case, but with a small $m_z^\ast$ that cannot be resolved experimentally (the ideal gas theory predicts $\approx 0.002$).
%The shift of $T_{c,1}$ is small, and could be an artifact due to the difficulty to detect small condensate fractions of a few percents. However, the shift of $T_{c,2}$ is too large to be explained by the same reason. 

Repulsive interactions between the atoms can be expected to lower the critical temperatures as in single-component gases \cite{giorgini1996a}, with an enhanced shift of $T_{\rm c,2}$ due to the presence of a condensate. %and to obtain a quantitative prediction for a spinor gas,  
We use a simplified version of Hartree-Fock (HF) theory to make quantitative predictions \cite{kawaguchi2012b}. Our self-consistent calculations include the trap potential in a semi-classical approximation, and treat the interactions as spin-independent. These approximations are valid only above $T_{\rm c,2}$, where at most one component condenses \cite{SM}. As a result, the HF model cannot make any prediction for the low-temperature behavior below $T_{\rm c,2}$. The results of the HF calculations, performed for atom numbers and trap frequencies matching the experimental values \cite{SM}, are shown in Figure\,\ref{fig3}. The HF model qualitatively accounts for the experimental data, explaining in particular the strong downwards shift of $T_{\rm c,2}$ for all $q$ and the shift of $m_z^\ast$ to higher values for $q/h\approx 8.9\,$kHz. %A noticeable feature for $q/h\approx 8.9\,$kHz is a ``jump'' of $T_{\rm c,2}$ when crossing $m_z^\ast$ from left to right, which is present in the calculation and in the data \cite{SM}. 
The residual discrepancy around $7-8$\,\% could be partially explained by finite-size and trap anharmonicity effects not included in the Hartree-Fock calculation \cite{SM}.

At the lowest field we studied, $q/h\approx 2.8\,$Hz (Fig.\,\ref{fig3}\,i-l), we observe a change in the \textit{nature} of $T_{\rm c,2}$. For high values of $m_z$, $T_{\rm c,2}$ corresponds to condensation into $m_F=-1$ while $m_F=0$ remains uncondensed. %
This phenomenon is incompatible with ideal gas theory \cite{isoshima2000a,lang2014a} \textit{and} with our HF model with spin-independent interactions. It corresponds to a change of the magnetic ordering appearing below $T_{\rm c,2}$. While coexisting $m_F=0$ and $m_F=+1$ components form a $M_\perp$ phase with $\langle \hat{S}_x+i\hat{S}_y \rangle \neq 0$, coexisting $m_F=\pm 1$ components correspond to a phase with $\langle \hat{S}_x+i\hat{S}_y \rangle = 0$ but where the spin-rotational symmetry around $z$ is broken by a non-zero spin-quadrupole tensor (``quasi-spin nematic'' phase -qSN). %This change in the magnetic properties of antiferromagnetic spinor condensates has been observed previously at very low temperatures \cite{liu2009a,jacob2012a,vinit2013a,jiang2014a,zhao2015a}.
At $T=0$ and in the single-mode regime, the $M_\perp$-\,qSN transition occurs at a critical magnetization $m_{z,c}=\sqrt{1-[1-(q/U_s)]^2}$, with $U_s \leq q$ the spin-dependent interaction energy \cite{zhang2003a}. When $q>U_s$, there is no phase transition and only the $M_\perp$ phase is present. This explains the qualitative difference between the data for $q/h=2.8\,$Hz and the other two values. We estimate $U_s/h \lesssim 50\,$Hz and $m_{z,\textrm{crit}} \approx 0.3$ for a BEC without thermal fraction \cite{jacob2012a}. This agrees well with the lowest temperature measurements reported in Fig.\,\ref{fig3}j-k. 

In the experimental data in Fig.\,\ref{fig3}\,i-l, the region of the phase diagram occupied by the $M_\perp$ phase shrinks with increasing temperature. In fact, we find that $m_F=-1$ condenses at $T_{c,2}$ for all parameters we have explored, with $m_F=0$ condensing at a third, lower critical temperature (except for $m_z \approx 0$, where all components appear to condense together within the accuracy of our measurement). 
%One can wonder whether the $M_\perp$-\,qSN transition follows from a mere extrapolation of the $T=0$ prediction with temperature-dependent $U_s$ and $m_{z}$ to account for the reduced condensed fraction. Such an extrapolation \cite{SM} appears in Figs.\,\ref{fig3}j-k as an almost vertical line, clearly not reproducing the measured boundary. We conjecture that spin coherences in the thermal components play a role at finite temperatures, as shown in \cite{kawaguchi2012b} for uniform systems. 
Finally, the dashed line in Fig.\,\ref{fig3}k shows $T_{\rm c,2}$ predicted by the HF model with spin-independent interactions. Although the model incorrectly predicts that $m_F=0$ should condense below $T_{\rm c,2}$, the predicted transition closely matches the observed boundary between single-component $m_F=+1$ BEC and qSN $m_F= \pm 1$ BEC. This indicates that the transition line itself (but not the magnetic order below it) is determined by the thermal component alone.
%However, our experimental results can be fully understood only when taking into account three important features, the harmonic trap (which is crucial to stabilize an antiferromagnetic condensate in a single spatial mode \cite{yi2002a}), the QZE and the interactions.

In conclusion, we have studied the finite-$T$ phase diagram of a spin-$1$ Bose gas with antiferromagnetic interactions. For condensates in the single-mode regime, we observed a sequence of transitions, two for high QZE and three for low QZE, with the lower two leading to different magnetic orders. We have found that a simplified HF model reproduces the trends observed in the variations of the critical temperatures $T_{\rm c,1}$ and $T_{\rm c,2}$ with magnetization and QZE. A more complete theoretical analysis accounting for all experimental features --in particular the harmonic trap, which is crucial to stabilize an antiferromagnetic condensate in a single spatial mode \cite{yi2002a}-- and elucidating the exact nature of the low-temperature transitions for low QZE remains open. A natural extension of this work would be to study the critical properties of the observed finite-$T$ transitions, in particular near $m_z=m_z^\ast$ and between the $M_\perp$ and qSN phases at very low $q$. Two-dimensional systems provide another intriguing direction to explore. Several Berezinskii-Kosterlitz-Thouless transitions mediated either by vortices or spin textures have been predicted \cite{mukerjee2006a,james2011a}. We expect that such topological features will further enrich the already complex phase diagram observed in three dimensions.

\begin{acknowledgments}
We acknowledge stimulating discussions with B. Evrard, L. De Sarlo, E. Witkowska, J. Beugnon, L. de Forges de Parny, A. Ran\c{c}on and T. Roscilde. This work has been supported by ERC (Synergy grant UQUAM). TZ acknowledges funding from the Hamburg Center for Ultrafast Imaging, and KJG from the European Union's Horizon 2020 research and innovation programme under the Marie {Sk\l{}odowska}-Curie grant agreement No. 701894.
\end{acknowledgments}
%
%\appendix

\bibliography{Tc_spinor_complet}

%\bibliographystyle{apsrev4-1}
%\begin{thebibliography}{37}
%\end{thebibliography}
%
%\begin{figure}
%\includegraphics[]{}
%\caption{} \label{}
%\end{figure}
%
\end{document}